\documentclass[12pt,letterpaper]{article}
\usepackage{amsmath,amssymb,pgf,pgfarrows,pgfnodes,float,appendix, hyperref}
\usepackage{graphicx}
\usepackage{subfigure}
\usepackage[margin=0.9in]{geometry}

\title{{\rm\footnotesize \qquad \qquad \qquad \qquad \qquad \ \qquad \qquad \qquad \ \ \ \ \ \                  RUNHETC-2017-11}\vskip.5in    Note on Localized Objects as Constrained States of Holographic Variables}
\author{Tom Banks\\
Department of Physics and NHETC\\
Rutgers University, Piscataway, NJ 08854\\
E-mail: \href{mailto:tibanks@ucsc.edu}{tibanks@ucsc.edu}}

\date{Oct. 1, 2017}
\begin{document}
\maketitle

\begin{abstract} We argue that localized excitations in Minkowski space must be thought of as constrained states of holographic degrees of freedom.  The Minkowski ``vacuum" is in fact a density matrix of infinite entropy.  The argument assumes that Minkowski space can be viewed as a limit of a space-time with non-vanishing cosmological constant, either positive or negative.  
\end{abstract}

\section{Introduction}

Discussions of models of quantum gravity in Minkowski space usually begin with the assumption that the observables of the model are matrix elements of a scattering operator $S$ in Fock space.  In particular, it is assumed that the vacuum is the only state with exactly zero energy.   We have known since the 1960s that this is incorrect in $4$ dimensions.  IR divergences in the presence of perturbative gravity demonstrate that all Fock space matrix elements of the S-operator vanish.  There have been a variety of proposals for constructing the correct Hilbert space, using partially resummed perturbation series\cite{kuletal}, following the ideas of Fadeev and Kulish\cite{fadkul} in QED.  

The situation in gravity is surely much more complicated and involves non-perturbative physics.  There are many kinematic regimes, including many where individual sub-energies $p_i \cdot p_j$ are all small, in which scattering of gravitons on each other produces systems of black holes which spiral around each other and merge and eventually decay to Hawking radiation.  The final quantum state of soft particles in such a collision is surely beyond the reach of any resummed perturbative calculation.  In \cite{soft} Fischler and I argued that these non-perturbative issues suggest strongly that even in high dimensions, the gravitational S-matrix is not unitary in Fock space, even though it has finite matrix elements there.  

This line of thought matched up with intuitions we had formed in our attempt to construct a general theory of quantum gravity based on the Covariant Entropy Principle\cite{BHGHtHJFSB}.  In these {\it Holographic Space-time} models, objects localized in causal diamonds always correspond to constrained states of a set of degrees of freedom living on the holographic screen of the 
diamond.  From about 2006\cite{refs} on, we realized that these degrees of freedom were connected to finite diamond deformations of the Awada-Gibbons-Shaw supersymmetric Bondi-Metzner-Sachs-van der Burg algebra, and thus to soft supergraviton states.

The purpose of the present note is to present evidence, independent of HST, for the idea that bulk-localized excitations are constrained states of boundary degrees of freedom.  The latter may be thought of as a fuzzy version of pure gauge modes of the (super)-gravitational field in the bulk of a causal diamond, which are non-trivial on the boundary of the diamond.  Here fuzzy means precisely that the expansion of functions in eigenmodes of the Dirac operator on the screen of the diamond is cut off, in a way that reflects the geometry of the diamond.  We will not present explicit evidence for the second part of this conjecture.

\section{de Sitter Space}

The idea of localized objects as constrained states was inspired by the formula for the metric of black holes in dS space
\begin{equation} ds^2 = - dt^2 (1 - \frac{R_S}{r} - \frac{r^2}{R^2}) + \frac{dr^2}{(1 - \frac{R_S}{r} - \frac{r^2}{R^2})} + r^2 d\Omega_2^2 . \end{equation}
The equation for horizons takes the form
\begin{equation} (r - R_+)(r - R_-)(r + R_+ + R_-) = 0, \end{equation}
where 
$$R_+ R_- (R_+ + R_-) = R_S R^2, $$ and
$$ R^2 = (R_+ + R_-)^2 - R_+ R_- = R_+^2 + R_-^2 + R_+ R_- . $$
The last equation shows that empty dS space has larger entropy than dS space with a local excitation. 
The entropy deficit is $- 2\pi R M$, as we expect for the thermal ensemble at the Gibbons-Hawking\cite{gh} temperature.  Note that this computation does not use quantum field theory, though $\hbar$ enters implicitly through the relation between area and entropy via the definition of the Planck area.  The covariant entropy principle suggests that the Hilbert space is finite dimensional\cite{tbwf}.  The entropy of empty dS space is, in this interpretation, the logarithm of the dimension of that Hilbert space, and one {\it defines energy} as the coefficient of the inverse temperature in the entropy deficit.  The entropy deficit itself is to be calculated from a microscopic model, which explains what the fundamental degrees of freedom are, and how one defines the constraints that correspond to bulk localization.  HST provides such a model\cite{bfm}.

 The Gibbons-Hawking argument can be viewed as a confirmation of this conjecture, which shows that if we put a quantum field in contact with the dS vacuum, it gets heated up to the dS temperature.  The fact that empty dS space acts as a heat bath for QFT, suggests strongly that the dS entropy is not to be associated with the degrees of freedom described by the QFT, but with a system of much larger entropy.  Indeed, in the context of the semi-classical expansion, the area contribution to entropy is always a leading order effect, and the entropy of field theoretic fluctuations around a classical background subleading.  It's only because of the divergence of the entropy of ``low energy" field theory degrees of freedom, which is a combination of the usual UV infinities of field theory with the infinite redshift of energies as one approaches a horizon, that one can try to model the classical entropy as a quantum correction.  Instead one should view this as a breakdown of effective field theory as far as the entropy calculation is concerned.  The firewall ``paradox"\cite{firewall} is clear evidence that the entropy of horizons cannot be explained, even approximately, by conventional QFT\footnote{Actually, if we view the horizon as a boundary of space-time, then we can consider pure gauge transformations of supergravity, which act non-trivially on the boundary, as candidates for the degrees of freedom describing the entropy.  This still gives an infinite overcounting of states, but in HST, a fuzzy version of this is in fact correct.}.  The firewall paradox is resolved by HST\cite{firewall3}.

However, our aim here is to argue for the interpretation of localized states as constrained states of horizon degrees of freedom, without resort to HST.
We note first that, in the thermodynamic approximation, the unconstrained vacuum ensemble is exactly degenerate.  This is inconsistent with the statement that it's a thermalized system, but of course a result in the thermodynamic approximation can be off by terms inversely proportional to powers of the entropy.  Indeed, thinking of the horizon states as located on a timelike stretched horizon a distance of order Planck scale from the true horizon, we expect them to have energies of order $1/R$, which would lead to thermalization if the spectrum is chaotic.  

There has been considerable skepticism about the idea that dS space has a finite dimensional Hilbert space.  The preferred alternative seems to be the radically infinite dimensional space suggested by quantum field theory in the dS background.  The main point of the present note, as far as dS space is concerned, is that none of this matters to any observation done within a single cosmological Horizon. Generic states entangling the finite causal patch with transhorizon degrees of freedom predict a maximally uncertain density matrix for the patch Hilbert space, if we believe the full Hilbert space has more than twice the entropy\footnote{When we use entropy in connection with a Hilbert space in this note, the reference is always to the entropy of the maximally uncertain density matrix.} of the patch Hilbert space.   Unless the patch Hilbert space entropy is just the Gibbons Hawking entropy, this would imply that the state of the system is highly unlikely to give the usual Gibbons Hawking temperature or have the Gibbons Hawking entropy.   This is just Page's theorem\cite{page}.   If the patch Hilbert space really has the Gibbons Hawking entropy, and empty dS space is in fact represented by an ensemble with that entropy, then the existence of the larger Hilbert space is irrelevant, in the sense that it is equivalent to the statement that the state of the patch Hilbert space corresponding to empty dS space is the maximally uncertain density matrix, and has no further consequences.  The same conclusion is suggested by the fact that causal patches associated with other geodesics are just given by active coordinate transformations on the dS patch. Since dS space has no time-like or null boundaries on which to define asymptotic symmetries which could act as unitary transformations on the physical Hilbert space, this just says that they are gauge copies of a given patch\cite{tbold}.  

Turning now to the Minkowski limit of vanishing c.c., we find an infinite dimensional Hilbert space of zero energy boundary states.   Localized states are constrained states with a large entropy deficit compared to the vacuum ensemble, and we must understand how to construct a scattering operator for those states.

\section{Anti de Sitter Space}

Now let us take the limit of Minkowski space from the opposite direction in c.c. space.  Here we have the advantage of a relatively well understood class of quantum models, which go under the rubric {\it Conformal Field Theories with a large radius dual}.  There is not complete agreement on the question of whether these models become exactly supersymmetric in the infinite radius limit, or whether the large radius AdS space is always accompanied by at least two extra large compact directions.  There is also a fairly well developed technology for obtaining Fock space matrix elements of the S matrix in $5$ or more Minkowski dimensions from CFT correlators, although Giddings has advanced arguments that this limit is more subtle than it appears\cite{gidd}.  

There has been a lot of discussion of entanglement of subsystems of the boundary QFT and its relation to the bulk geometry.  However, the regions under discussion are always infinite, and touch the boundary regions that define subsystems of the boundary CFT.  Scale radius duality in AdS/CFT leads one to surmise that the description of finite regions must involve elimination of UV degrees of freedom in the CFT. The physically motivated discussions of the limit of boundary correlators which give S-matrix elements (as opposed to the purely mathematical calculations involving Mellin space) always emphasized a finite causal diamond, called ``the arena" , in which the geometry of AdS was almost indistinguishable from that of Minkowski space.  One is thus motivated to ask what the state of the subsystem corresponding to ``the Arena" is in some finite energy state of the boundary QFT.  Note that there is exactly one time-like geodesic in AdS space, such that the Arena's causal diamond is the causal diamond of an interval of sufficiently small proper time along that geodesic.  Correspondingly, there is a particular choice of generator $K_0 + P_0$ among the conjugacy class of generators of the conformal group whose commutant is isomorphic $SO(d - 1)$, which corresponds to the Arena.

We can immediately see that this is incompatible with the naive expectation that the Minkowski ground state is the limit of the AdS ground state and the Minkowski Hilbert space is obtained by acting with simple operators on that ground state, in a Fock space like picture.  This follows from the simple assumption that the Hilbert space of the Arena has finite dimension, with entropy $\ll R^{d-2}$ where $R$ is the AdS radius in $d$ dimensional Planck units and $d$ is the dimension of the Minkowski space that emerges in the large radius limit. 

The CFT Hilbert space has infinite dimension, and even if we cut it off by considering\footnote{invoking scale radius duality} a field theory corresponding to a finite area causal diamond with radius $\gg R$ its dimension is exponentially larger than that of the subsystem corresponding to the Arena.
Therefore, by Page's Theorem\cite{page}, in a typical state in the full Hilbert space, the Arena is maximally entangled with a much larger system, and its reduced density matrix is maximally uncertain.  This is consistent with the physical picture of the Arena as a primarily empty almost Minkowski space-time, only if we assume that that space-time does not correspond to a pure state, but rather a maximally uncertain density matrix. 

One could respond to this by claiming that the ground state of the QFT is, at least approximately, a tensor product between a unique state of the Arena subsystem and a state of its complement in the (cutoff) QFT Hilbert space\footnote{The notion of tensor factorization is much more complex for infinite dimensional Hilbert spaces.}.   The problem with this idea is that the QFT ground state is unique.  In the case of a CFT different versions of $K_0 + P_0$ in the conjugacy class all share the same ground state, but they correspond to different time-like geodesics, which are not at rest relative to each other.  That is, the Arena for one choice of $K_0 + P_0$ is not the same space-time region as that for another.  

In fact, there's a problem even for a fixed choice of $K_0 + P_0$.
There is, as we have said, a correspondence between a choice of $K_0 + P_0$ and a timelike trajectory, but that correspondence does not specify an interval along that trajectory.  If we consider a small excitation of the QFT vacuum, designed to send a small number of particles into the Arena, to undergo a scattering event that approximates Minkowski space scattering, among the criteria for the validity of that approximation is that the time elapsed in that event\cite{In scattering theory, notions like time delays, and the time elapsed in an event, are encoded in the energy dependence of the eigenvalues of the S-matrix.} be much shorter than the AdS radius.  For example, to have a hope of approximating black hole formation and evaporation in Minkowski space both the formation time and evaporation time must be shorter than the AdS radius.  On longer time scales, the final state particles bounce off the conformal boundary and are very unlikely to enter again the diamond which is simply the time translation of the Arena.   The empty Arena itself could be the time translation of a diamond where a scattering process took place a time $t > R$ in the past.  Thus, it's clear that the approximate Minkowski vacuum in the Arena can be achieved in {\it an infinite number of states of the QFT}.  It's inconsistent to claim that all of these states are well approximated by a tensor product of a unique state of the Arena times different states of the DOF that commute with the Arena variables.  In particular, consider variables in the Arena, which evolve into variables of the time translated Arena, under the action of the QFT Hamiltonian.  If a scattering process took place in the Arena, these degrees of freedom are definitely entangled with degrees of freedom that are no longer localized in the translated diamond.  Removing that entanglement would require dramatic violations of bulk locality.  The state of the Arena corresponding to the Minkowski vacuum is thus impure, generically, and given the large size of the system with which the Arena is entangled, the density matrix is maximally uncertain.  

Parenthetically, we remark that the restriction of an AdS mimicry of a Minkowski process to a proper time along a geodesic in the Arena that is $< R$, shows us that, even at the perturbative level, one cannot obtain complete information about the Minkowski S matrix from AdS correlators.  Consider the formation and evaporation of a black hole of Schwarzschild radius $R_S \ll R$.
In $d$ dimensions, in order to capture the quantum mechanics of the complete process, the Page time $ \sim R_S^{d - 1}$ must be $\ll R$. Now consider a Hawking particle, emitted in the late stages of the evaporation.  In Minkowski space, there are finite amplitudes for that particle to emit gravitons at arbitrarily late times, and the full S-matrix includes those amplitudes.  CFT correlators at such late times capture the effect of particles bouncing off the wall, and will not coincide with the corresponding Minkowski matrix elements.  Note that this description was completely covariant.  The rest frame and position of the black hole define a time-like geodesic, and our use of time in the above description refers to the proper time along that geodesic.  A finite time interval along that geodesic defines the Arena, and the Minkowski S matrix includes processes that take place outside the Minkowski diamond, which approximately coincides with the Arena.  The remarks of this paragraph appear to be related to issues raised by Giddings in \cite{gidd}.

Returning to the description of the Arena, the fact that the typical state corresponding to the Minkowski vacuum has entropy of order the Arena's area shows us that the Minkowski Hamiltonian {\it cannot} coincide with that of the boundary QFT.  The latter is gapped and does not have an exponentially large number of low energy states.  In order to be consistent with physics in Minkowski space, the actual Hamiltonian describing evolution in the proper time of the time-like geodesic connecting the past and future tips of the Arena, must have a huge set of almost degenerate low energy states, whose energy goes to zero as the AdS radius goes to infinity.  

Once we've accepted that the Minkowski vacuum constructed from a limit of AdS systems is a large entropy density matrix, we are forced to the conclusion that localized excitations inside the Arena must correspond to low entropy constrained states of the degrees of freedom responsible for the Arena's entropy.  This is the same lesson we learned from the black hole entropy formula in dS space, but that formula gives us a more precise idea of how the number of constraints is related to the size of the Arena causal diamond and the eigenvalue of the Minkowski Hamiltonian.  

\section{Conclusions}

Simple considerations involving only the general properties of entropy and entanglement lead to the conclusion that if we try to construct a quantum theory of Minkowski space-time, by taking the limit of vanishing c.c. from either positive or negative directions, the Minkowski vacuum is a maximally uncertain density matrix, on a Hilbert space whose dimension goes to infinity exponentially with a power of the c.c. .  Bulk localized excitations must correspond to constrained states of the degrees of freedom responsible for the entropy.

We've also added to the evidence that computation of CFT correlators cannot lead, in the limit $R_{AdS}/L_P \rightarrow\infty$ to the full Minkowski S-matrix.
Previous arguments we've advanced to that effect\cite{hstads}\cite{soft} had to do with finite time behavior of black holes or S-matrix elements in non-Fock sectors of the Hilbert space.  Here we've pointed out the existence of perturbative processes, which cannot be reproduced for any finite value of $R$.  More importantly, we've argued that the density matrix representing empty Minkowski space in AdS/CFT, is highly impure and cannot be identified with the unique CFT vacuum.

In this note, we've deliberately steered away from invoking the Holographic Space Time formalism (HST), but it's worth pointing out how HST circumvents all of these difficulties, and gives us a universal picture of maximally symmetric quantum space-times.  HST starts from the quantum theory of a finite interval of proper time along a time-like geodesic in such a space-time\footnote{In HST the idea that the metric has large quantum fluctuations is not correct. The metric is a hydrodynamic field describing dimensions of Hilbert spaces in the quantum theory.  Only its small fluctuations around a low entropy state can be viewed, approximately, as quantum fields.}.  In time symmetric space-times the evolution is most conveniently formulated in terms of causal diamonds for symmetric intervals $[t, - t]$.  The variables are spinors on the $d - 2$ dimensional holographic screens of these diamonds, with an angular momentum cutoff related to the area of the screen, and the time dependent Hamiltonian couples together only those variables associated with the diamond at any time.  One makes $d - 2$ forms out of spinors, so that the Hamiltonian is constructed to be invariant under the fuzzy version of volume preserving maps.  This leads to fast scrambling on the screen, which can account for the behavior of Minkowski black holes.  In the limit of an infinite Minkowksi like screen, with screen area growing like $t^{d-2}$ the spinor variables converge to currents, which carry the quantum numbers of stable particles through and on the screen.   The currents are operator valued measures on the momentum null cone and satisfy a generalization of the AGS\cite{agstb} SUSic version of the BMSV\cite{bmsv} algebra.  The momentum null cone can be thought of as the spectrum of the BMSV sub-algebra.  

Particle-like states, with infinitely concentrated currents, are rare.  The generic scattering state has jets of particles instead, but most of its entropy comes from soft states concentrated near the tip of the null cone $P = 0$.  Jets are defined by constraints isolating them from the soft states: the $ P = 0$ currents, which are still functions of the angles on the sphere, must vanish in small annuli surrounding the opening angle of each jet.  The formulation of the theory in finite causal diamonds allows for a very precise definition of this jet isolation criterion.

The modifications of this picture for non-zero c.c. are simple, particularly for positive c.c. .  In that case we simply stop the growth of the Hilbert space at a fixed finite dimension, determined by the Gibbons-Hawking entropy formula.  The time dependent Hamiltonian converges to a time independent one, which is still a fast scrambler.  The energies of typical states scale like $1/N$ in Planck units, where $NL_P$ is the dS radius.
This accounts for the behavior of the dS horizon when it is perturbed.  The jet isolation of the Minkowski theory becomes the definition of states with energies of order $1$, rather than order $1/N$, which can be (temporarily) localized inside the bulk of the dS diamond\footnote{The different scaling of localized vs. boundary energies is simply related to the redshift of the time translation operator in the static coordinate system.}.  These are states satisfying of order $N$ constraints, and the (statistical plus quantum) probability of finding such a state is of order $e^{ - E N}$  .  $E$ is proportional to the energy in Planck units.  Thus this formalism explains the temperature of dS space.  

AdS space is a bit more complex because the area of causal diamonds goes to infinity at a finite time.  We know that the limit is controlled in the standard Wilsonian way, by a CFT on the holographic screen at infinity.  Volume preserving mapping invariance is incompatible with the properties of field theory, which require a fixed conformal structure\footnote{We are here speaking of global rather than gauged volume preserving mapping symmetries.}.  To achieve a field theoretic limit we must change the rule for the time evolution of the Hamiltonian, which we used in Minkowski and dS space, and replace it by an inverse tensor network renormalization group (TNRG) flow\cite{Evenblyvidaltbwftb}, at a time when the causal diamond has an area of order the AdS radius.  The Minkowski dynamics on scales much less than
the AdS radius is ``forgotten" by the TNRG dynamics, which converges to the field theory.  {\it HST gives us an explicit finite mapping between bulk dynamics and boundary CFT, in which locality on scales much smaller than the AdS radius is not accessible to any correlation function in the CFT.} Furthermore, although the proposed flat space dynamics is not yet correct, since it has not incorporated the requirement that information accessible to trajectories moving at relative velocity is described in a compatible manner, it provides a large class of candidate Hamiltonians, all of which incorporate unitarity and locality, and exhibit excitations with all of the qualitative characteristics of both black holes and jets of elementary particles.

\vskip.3in
\begin{center}
{\bf Acknowledgments }\\
The work of T.Banks is {\bf\it NOT} supported by the Department of Energy, the NSF, the Simons Foundation, the Templeton Foundation or FQXI. \end{center}


\begin{thebibliography}{99}
\bibitem{tbas} T.~Banks and W.~Fischler, ``Holographic Space-time, Newton's Law and the Dynamics of Black Holes,'' arXiv:1606.01267 [hep-th];
   T.~Banks, ``Current Algebra on the Conformal Boundary and the Variables of Quantum Gravity,''
  arXiv:1511.01147 [hep-th];
   T.~Banks, ``The Temperature/Entropy Connection for Horizons, Massless Particle Scattering, and the Origin of Locality,'' Int.\ J.\ Mod.\ Phys.\ D {\bf 24}, no. 12, 1544010 (2015)
  doi:10.1142/S0218271815440101  [arXiv:1505.04273 [hep-th]];
  T.~Banks,``The Super BMS Algebra, Scattering and Holography,''
  arXiv:1403.3420 [hep-th];
 \bibitem{fadkul}   P.~P.~Kulish and L.~D.~Faddeev,
  ``Asymptotic conditions and infrared divergences in quantum electrodynamics,''
  Theor.\ Math.\ Phys.\  {\bf 4}, 745 (1970)
  [Teor.\ Mat.\ Fiz.\  {\bf 4}, 153 (1970)];
  doi:10.1007/BF01066485
 \bibitem{kuletal} ``Asymptotical states of massive particles interacting with gravitational field,''
  Teor.\ Mat.\ Fiz.\  {\bf 6}, 28 (1971);
  doi:10.1007/BF01037574
   P.P. Kulish "Infrared divergences of quantized
  gravitational field", Zapiski Nauchnykh Seminarov
  LOMI, vol.77, 106 - 123 (1978); J.~Ware, R.~Saotome and R.~Akhoury, ``Construction of an asymptotic S matrix for perturbative quantum gravity,'' JHEP {\bf 1310}, 159 (2013) doi:10.1007/JHEP10(2013)159
  [arXiv:1308.6285 [hep-th]];  R.~Akhoury, R.~Saotome and G.~Sterman, ``Collinear and Soft Divergences in Perturbative Quantum Gravity,'' Phys.\ Rev.\ D {\bf 84}, 104040 (2011)
  doi:10.1103/PhysRevD.84.104040 [arXiv:1109.0270 [hep-th]];
   W.~Donnelly and S.~B.~Giddings, ``Observables, gravitational dressing, and obstructions to locality and subsystems,'' Phys.\ Rev.\ D {\bf 94}, no. 10, 104038 (2016) [arXiv:1607.01025 [hep-th]];
  \bibitem{soft} T.~Banks and W.~Fischler, ``Soft Gravitons and the Flat Space Limit of Anti-deSitter Space,'' arXiv:1611.05906 [hep-th].
  \bibitem{BHGHtHJFSB}  J.~D.~Bekenstein,
  ``Black holes and entropy,''
  Phys.\ Rev.\ D {\bf 7}, 2333 (1973),
  doi:10.1103/PhysRevD.7.2333;
   J.~D.~Bekenstein,
  ``Generalized second law of thermodynamics in black hole physics,''
  Phys.\ Rev.\ D {\bf 9}, 3292 (1974).
  doi:10.1103/PhysRevD.9.3292
  S.~W.~Hawking,
  ``Particle Creation by Black Holes,''
  Commun.\ Math.\ Phys.\  {\bf 43}, 199 (1975)
  Erratum: [Commun.\ Math.\ Phys.\  {\bf 46}, 206 (1976)];
  doi:10.1007/BF02345020
   S.~W.~Hawking,
  ``Black Holes and Thermodynamics,''
  Phys.\ Rev.\ D {\bf 13}, 191 (1976),
  doi:10.1103/PhysRevD.13.191;
   S.~W.~Hawking,
  ``Breakdown of Predictability in Gravitational Collapse,''
  Phys.\ Rev.\ D {\bf 14}, 2460 (1976),
  doi:10.1103/PhysRevD.14.2460;
   G.~W.~Gibbons and S.~W.~Hawking,
  ``Cosmological Event Horizons, Thermodynamics, and Particle Creation,''
  Phys.\ Rev.\ D {\bf 15}, 2738 (1977),
  doi:10.1103/PhysRevD.15.2738;
  G.~'t Hooft,
  ``The Black hole horizon as a quantum surface,''
  Phys.\ Scripta T {\bf 36}, 247 (1991),
  doi:10.1088/0031-8949/1991/T36/026;
   T.~Jacobson, ``Thermodynamics of space-time: The Einstein equation of state,''
  Phys.\ Rev.\ Lett.\  {\bf 75}, 1260 (1995)
  doi:10.1103/PhysRevLett.75.1260
  [gr-qc/9504004];
  W.~Fischler and L.~Susskind, ``Holography and cosmology,''
  hep-th/9806039.
  R.~Bousso, ``A Covariant entropy conjecture,''
  JHEP {\bf 9907}, 004 (1999)
  doi:10.1088/1126-6708/1999/07/004
  [hep-th/9905177];
   R.~Bousso,``Holography in general space-times,''
  JHEP {\bf 9906}, 028 (1999)
  doi:10.1088/1126-6708/1999/06/028
  [hep-th/9906022].
  \bibitem{refs}  T.~Banks, ``II(infinity) Factors and M-theory in Asymptotically Flat Space-Time,''  hep-th/0607007;
  T.~Banks and W.~Fischler,``Holographic Space-time and Newton's Law,''
  arXiv:1310.6052 [hep-th];
   T.~Banks,
  ``Lectures on Holographic Space Time,''
  arXiv:1311.0755 [hep-th];
T.~Banks, W.~Fischler, S.~Kundu and J.~F.~Pedraza,
  ``Holographic Space-time and Black Holes: Mirages As Alternate Reality,''
  arXiv:1401.3341 [hep-th];
   T.~Banks,
  ``Supersymmetry Breaking and the Cosmological Constant,''
  Int.\ J.\ Mod.\ Phys.\ A {\bf 29}, 1430010 (2014)
  doi:10.1142/S0217751X14300105
  [arXiv:1402.0828 [hep-th]];
  T.~Banks,
  ``The Super BMS Algebra, Scattering and Holography,''
  arXiv:1403.3420 [hep-th];
  T.~Banks,
  ``The Temperature/Entropy Connection for Horizons, Massless Particle Scattering, and the Origin of Locality,''
  Int.\ J.\ Mod.\ Phys.\ D {\bf 24}, no. 12, 1544010 (2015)
  doi:10.1142/S0218271815440101
  [arXiv:1505.04273 [hep-th]];
   T.~Banks,
  ``Current Algebra on the Conformal Boundary and the Variables of Quantum Gravity,''
  arXiv:1511.01147 [hep-th];
  T.~Banks and W.~Fischler,
  ``Holographic Space-time, Newton's Law and the Dynamics of Black Holes,''
  arXiv:1606.01267 [hep-th].
  \bibitem{gh}  G.~W.~Gibbons and S.~W.~Hawking,
  ``Cosmological Event Horizons, Thermodynamics, and Particle Creation,''
  Phys.\ Rev.\ D {\bf 15}, 2738 (1977),
  doi:10.1103/PhysRevD.15.2738;

  \bibitem{tbwf} W.~Fischler, ``Taking de Sitter Seriously", W. Fischler, ÒTaking de Sitter Seriously.Ó Talk given at Role of Scaling Laws in Physics and Biology (Celebrating the 60th Birthday of Geoffrey West), Santa Fe, Dec., 2000. ; T.~Banks, Lecture at the Lennyfest, Stanford University, May 2000; T.~Banks,
  ``Cosmological breaking of supersymmetry?,''
  Int.\ J.\ Mod.\ Phys.\ A {\bf 16}, 910 (2001)
  doi:10.1142/S0217751X01003998
  [hep-th/0007146].
\bibitem{hstads} T.~Banks and W.~Fischler, ``Holographic Space-time Models of Anti-deSitter Space-times,'' arXiv:1607.03510 [hep-th]; T.~Banks and W.~Fischler,
  \bibitem{firewall} 	Black Holes: Complementarity or Firewalls? - Almheiri, Ahmed et al. JHEP 1302 (2013) 062 arXiv:1207.3123 [hep-th];
Better Late than Never: Information Retrieval from Black Holes - Braunstein, Samuel L. et al. Phys.Rev.Lett. 110 (2013) no.10, 101301 arXiv:0907.1190 [quant-ph];
Better Late than Never: Information Retrieval from Black Holes - Braunstein, Samuel L. et al. Phys.Rev.Lett. 110 (2013) no.10, 101301 arXiv:0907.1190 [quant-ph];
Complementarity Is Not Enough - Bousso, Raphael Phys.Rev. D87 (2013) no.12, 124023 arXiv:1207.5192 [hep-th];
Complementarity Endures: No Firewall for an Infalling Observer - Nomura, Yasunori et al. JHEP 1303 (2013) 059 arXiv:1207.6626 [hep-th] UCB-PTH-12-12;
Comments on black holes I: The possibility of complementarity - Mathur, Samir D. et al. JHEP 1401 (2014) 034 arXiv:1208.2005 [hep-th];
Is Alice burning or fuzzing? - Chowdhury, Borun D. et al. Phys.Rev. D88 (2013) 063509 arXiv:1208.2026 [hep-th] IPHT-T12-063;
Singularities, Firewalls, and Complementarity - Susskind, Leonard arXiv:1208.3445 [hep-th];
Holographic Space-Time Does Not Predict Firewalls - Banks, Tom et al. arXiv:1208.4757 [hep-th] UTTG-15-12, TCC-015-12, RUNHETC-2012-17, SCIPP-12-11;
Firewall or smooth horizon? - Ori, Amos Gen.Rel.Grav. 48 (2016) no.1, 9 arXiv:1208.6480 [gr-qc];
Non-extremal Black Hole Microstates: Fuzzballs of Fire or Fuzzballs of Fuzz ? - Bena, Iosif et al. JHEP 1212 (2012) 014 arXiv:1208.3468 [hep-th];
String Theory Versus Black Hole Complementarity - Giveon, Amit et al. JHEP 1212 (2012) 094 arXiv:1208.3930 [hep-th];
Origin of the blackhole information paradox - Brustein, Ram Fortsch.Phys. 62 (2014) 255-265 arXiv:1209.2686 [hep-th];
The Transfer of Entanglement: The Case for Firewalls - Susskind, Leonard arXiv:1210.2098 [hep-th];
Comment on the black hole firewall - Hossenfelder, Sabine arXiv:1210.5317 [gr-qc];
Unitarity and fuzzball complementarity: 'Alice fuzzes but may not even know it!' - Avery, Steven G. et al. JHEP 1309 (2013) 012 arXiv:1210.6996 [hep-th];
Black Holes, Information, and Hilbert Space for Quantum Gravity - Nomura, Yasunori et al. Phys.Rev. D87 (2013) 084050 arXiv:1210.6348 [hep-th] MIT-CTP-4405, UCB-PTH-12-17;
Is the firewall consistent?: Gedanken experiments on black hole complementarity and firewall proposal - Hwang, Dong-il et al. JCAP 1301 (2013) 005 arXiv:1210.6733 [gr-qc];
Remarks on Black Hole Evolution a la Firewalls and Fuzzballs - Rama, S.Kalyana arXiv:1211.5645 [hep-th] IMSC-2012-11-18;
A Note on (No) Firewalls: The Entropy Argument - Nomura, Yasunori et al. JHEP 1307 (2013) 124 arXiv:1211.7033 [hep-th] MIT-CTP-4425, UCB-PTH-12-19;
Nonviolent nonlocality - Giddings, Steven B. Phys.Rev. D88 (2013) 064023 arXiv:1211.7070 [hep-th];
Black holes without firewalls - Larjo, Klaus et al. Phys.Rev. D87 (2013) no.10, 104018 arXiv:1211.4620 [hep-th] BROWN-HET-1636, NORDITA-2012-90, RH-10-2012;
Empty black holes, firewalls, and the origin of BekensteinÐHawking entropy - Saravani, Mehdi et al. Int.J.Mod.Phys. D23 (2015) no.13, 1443007 arXiv:1212.4176 [hep-th];
Boundary unitarity and the black hole information paradox - Jacobson, Ted Int.J.Mod.Phys. D22 (2013) 1342002 arXiv:1212.6944 [hep-th];
Black Hole Complementarity and the Harlow-Hayden Conjecture - Susskind, Leonard arXiv:1301.4505 [hep-th];
Quantum Computation vs. Firewalls - Harlow, Daniel et al. JHEP 1306 (2013) 085 arXiv:1301.4504 [hep-th];
Black hole complementarity and firewall in two dimensions - Kim, Wontae et al. JHEP 1305 (2013) 060 arXiv:1301.5138 [gr-qc] YITP-13-33;
Macroscopic superpositions and black hole unitarity - Hsu, Stephen D.H. arXiv:1302.0451 [hep-th];
Nonviolent information transfer from black holes: A field theory parametrization - Giddings, Steven B. Phys.Rev. D88 (2013) no.2, 024018 arXiv:1302.2613 [hep-th];
A Self-consistent Model of the Black Hole Evaporation - Kawai, Hikaru et al. Int.J.Mod.Phys. A28 (2013) 1350050 arXiv:1302.4733 [hep-th] KUNS-2434, KEK-TH-1605;
Status report: black hole complementarity controversy - Lee, Bum-Hoon et al. Nucl.Phys.Proc.Suppl. 246-247 (2014) 178-182 arXiv:1302.6006 [gr-qc]
No Firewalls in Holographic Space-Time or Matrix Theory - Banks, T. et al. arXiv:1305.3923 [hep-th];
Pure states and black hole complementarity - Lowe, David A. et al. Phys.Rev. D88 (2013) 044012 arXiv:1305.7459 [hep-th] BROWN-HET-1644, NORDITA-2013-36;
Excluding Black Hole Firewalls with Extreme Cosmic Censorship - Page, Don N. JCAP 1406 (2014) 051 arXiv:1306.0562 [hep-th] ALBERTA-THY-4-13
[3];	D. Marolf. The case for firewalls - talk at KITP workshop, Aug 19-30, 2013. http://online.kitp.ucsb.edu/online/fuzzorfirem13/marolf/
   \bibitem{firewall3} T.~Banks, W.~Fischler, S.~Kundu and J.~F.~Pedraza,
  ``Holographic Space-time and Black Holes: Mirages As Alternate Reality,''
  arXiv:1401.3341 [hep-th].
  \bibitem{page} D.~N.~Page, ``Average entropy of a subsystem,''
  Phys.\ Rev.\ Lett.\  {\bf 71}, 1291 (1993)
  doi:10.1103/PhysRevLett.71.1291
  [gr-qc/9305007].
  \bibitem{tbold}  T.~Banks,``Some thoughts on the quantum theory of stable de Sitter space,''
  hep-th/0503066.
  \bibitem{agstb}  M.~A.~Awada, G.~W.~Gibbons and W.~T.~Shaw, ``Conformal Supergravity, Twistors And The Super Bms Group,'' Annals Phys.\  {\bf 171}, 52 (1986);
  doi:10.1016/S0003-4916(86)80023-9
  T.~Banks, ``II(infinity) Factors and M-theory in Asymptotically Flat Space-Time,''  hep-th/0607007;
T.~Banks,
  ``The Super BMS Algebra, Scattering and Holography,''
  arXiv:1403.3420 [hep-th].
\bibitem{Evenblyvidaltbwftb}	G. Evenbly and G. Vidal. 2015. arXiv:cond-matt/. Phys.Rev.Lett.,115,180405 arXiv:1412.0732;
G. Evenbly, G. Vidal. 2015. [12] arXiv:condmatt/. Phys.Rev.Lett.,115,200401 arXiv:1502.05385;
G. Evenbly. Algorithms for tensor network renormaliza- tion arXiv:condmatt/. arXiv:1509.07484;
G. Evenbly, P. Corboz, G. Vidal. 2010. arXiv:cond-matt/. Phys.Rev.,B82,132411 arXiv:0912.2166;
T.~Banks and W.~Fischler,
  ``Holographic Space-time Models of Anti-deSitter Space-times,''
  arXiv:1607.03510 [hep-th];
  T.~Banks, ``Tensor Network Renormalization Group for General Conformal Field Theories and Holographic Space-time Models for Anti de Sitter Space", {\it manuscript in preparation}.
  \bibitem{esw}G.~F.~Sterman and S.~Weinberg, ``Jets from Quantum Chromodynamics,''
  Phys.\ Rev.\ Lett.\  {\bf 39}, 1436 (1977). doi:10.1103/PhysRevLett.39.1436; T.~Banks, Kodosky Lectures, July 2016 U. Texas, Austin, see http://zippy.ph.utexas.edu/videos.html ; 
  T.~Banks and W.~Fischler, ``Holographic Space-time, Newton's Law and the Dynamics of Black Holes,'' arXiv:1606.01267 [hep-th].
  \bibitem{bfm}   T.~Banks, B.~Fiol and A.~Morisse,
  ``Towards a quantum theory of de Sitter space,''
  JHEP {\bf 0612}, 004 (2006)
  doi:10.1088/1126-6708/2006/12/004
  [hep-th/0609062].
  
  \bibitem{adscft}
  O.~Aharony, S.~S.~Gubser, J.~M.~Maldacena, H.~Ooguri and Y.~Oz,
  ``Large N field theories, string theory and gravity,''
  Phys.\ Rept.\  {\bf 323}, 183 (2000)
  doi:10.1016/S0370-1573(99)00083-6
  [hep-th/9905111].
  \bibitem{elitzurdavid} S.~Elitzur,``The Applicability of Perturbation Expansion to Two-dimensional Goldstone Systems,'' Nucl.\ Phys.\ B {\bf 212}, 501 (1983);
  doi:10.1016/0550-3213(83)90682-X
  F.~David, ``Cancellations of Infrared Divergences in the Two-dimensional Nonlinear Sigma Models,''
  Commun.\ Math.\ Phys.\  {\bf 81}, 149 (1981);
  doi:10.1007/BF01208892
  F.~David,``Quantization With a Global Constraint and Infrared Finiteness of Two-dimensional Goldstone Systems,'' Nucl.\ Phys.\ B {\bf 190}, 205 (1981).
  doi:10.1016/0550-3213(81)90490-9
  \bibitem{hp}  S.~W.~Hawking and D.~N.~Page,``Thermodynamics of Black Holes in anti-De Sitter Space,'' Commun.\ Math.\ Phys.\  {\bf 87}, 577 (1983).
  doi:10.1007/BF01208266
  \bibitem{shenkergrossmende}  D.~J.~Gross and P.~F.~Mende,``String Theory Beyond the Planck Scale,'' Nucl.\ Phys.\ B {\bf 303}, 407 (1988);
  doi:10.1016/0550-3213(88)90390-2
  D.~J.~Gross and P.~F.~Mende, ``The High-Energy Behavior of String Scattering Amplitudes,''
  Phys.\ Lett.\ B {\bf 197}, 129 (1987).
  doi:10.1016/0370-2693(87)90355-8
   S.~H.~Shenker, ``The Strength of nonperturbative effects in string theory,''
  In *Brezin, E. (ed.), Wadia, S.R. (ed.): The large N expansion in quantum field theory and statistical physics* 809-819
\bibitem{previous} T.~Banks and W.~Fischler,
  ``Holographic Inflation Revised,''
  arXiv:1501.01686 [hep-th];

T.~Banks,
  ``Holographic Space-Time: The Takeaway,''
  arXiv:1109.2435 [hep-th];
 T.~Banks,
  ``TASI Lectures on Holographic Space-Time, SUSY and Gravitational Effective Field Theory,''
  arXiv:1007.4001 [hep-th];
  \bibitem{holoinflation2} T.~Banks and W.~Fischler,
  ``Holographic Inflation Revised,''
  arXiv:1501.01686 [hep-th].
  \bibitem{holonewton2} 
  T.~Banks and W.~Fischler,
  ``Holographic Space-time, Newton's Law and the Dynamics of Black Holes,''
  arXiv:1606.01267 [hep-th]; 
  \bibitem{bmsv} H. Bondi. Gravitational waves in general relativity VII. Waves from isolated axisymmetric systems - 1962; M. G. J. van der Burg, A. W. K Metzner. Proc.Roy.Soc.Lond.,269,21
R. K. Sachs. Gravitational waves in general relativity VIII. Waves in asymptotically flat space-time - 1962. Proc.Roy.Soc.Lond.,270,103; A unified treatment of null and spatial infinity in general relativity. I - Universal structure, asymptotic symmetries, and conserved quantities at spatial infinity - Ashtekar, A. et al. J.Math.Phys. 19 (1978) 1542-1566
Asymptotic Quantization of the Gravitational Field - Ashtekar, A. Phys.Rev.Lett. 46 (1981) 573-576
A. Ashtekar and M. Streubel. Symplectic Geometry of Radiative Modes and Conserved Quantities at Null Infinity - 1981. Proc.Roy.Soc.Lond.,376,585
A. Ashtekar. Asymptotic Quantization: Based On 1984 Naples Lectures - Naples, Italy: Bibliopolis,(1987); T. Banks, A Critique of pure string theory: Heterodox opinions of diverse dimensions, hep-th/0306074; G. Barnich and C. Troessaert, BMS charge algebra, JHEP 1112, 105 (2011) [arXiv:1106.0213 [hep-th]]; G. Barnich and C. Troessaert, Supertranslations call for superrotations, PoS, 010 (2010) [Ann. U. Craiova Phys. 21, S11 (2011)] [arXiv:1102.4632 [gr-qc]]; G. Barnich and C. Troessaert, Symmetries of asymptotically flat 4 dimensional spacetimes at null infinity revisited, Phys. Rev. Lett. 105, 111103 (2010) [arXiv:0909.2617 [gr-qc]]; D. Kapec, V. Lysov, S. Pasterski and A. Strominger, Semiclassical Virasoro symmetry of the quantum gravity S-matrix, JHEP 1408, 058 (2014) [arXiv:1406.3312 [hep-th]]; A. Strominger, On BMS Invariance of Gravitational Scattering, JHEP 1407, 152 (2014) [arXiv:1312.2229 [hep-th]]. T. He, V. Lysov, P. Mitra and A. Strominger, BMS supertranslations and Weinbergs soft graviton theorem, JHEP 1505, 151 (2015) [arXiv:1401.7026 [hep-th]]; F. Cachazo and A. Strominger, Evidence for a New Soft Graviton Theorem, arXiv:1404.4091 [hep-th]; T. He, P. Mitra, A. P. Porfyriadis and A. Strominger, New Symmetries of Massless QED, JHEP 1410, 112 (2014) [arXiv:1407.3789 [hep-th]]; A. Strominger and A. Zhiboedov, Gravitational Mem- ory, BMS Supertranslations and Soft Theorems, arXiv:1411.5745 [hep-th]. D. Kapec, V. Lysov and A. Strominger, Asymptotic Symmetries of Massless QED in Even Di- mensions, arXiv:1412.2763 [hep-th]. D. Kapec, V. Lysov, S. Pasterski and A. Strorefsdelminger, Higher-Dimensional Supertranslations and WeinbergÕs Soft Graviton Theorem, arXiv:1502.07644 [gr-qc];   B.~Gabai and A.~Sever,``Large Gauge Symmetries and Asymptotic States in QED,'' arXiv:1607.08599 [hep-th];
   M.~Mirbabayi and M.~Porrati, ``Shaving off Black Hole Soft Hair,''
  arXiv:1607.03120 [hep-th];
   C.~Gomez and M.~Panchenko,
  ``Asymptotic dynamics, large gauge transformations and infrared symmetries,''
  arXiv:1608.05630 [hep-th];
   A. Strominger, On BMS Invariance of Gravitational Scattering, JHEP 1407, 152 (2014) [arXiv:1312.2229 [hep-th]]. T. He, V. Lysov, P. Mitra and A. Strominger, BMS supertranslations and Weinbergs soft graviton theorem, JHEP 1505, 151 (2015) [arXiv:1401.7026 [hep-th]]; F. Cachazo and A. Strominger, Evidence for a New Soft Graviton Theorem, arXiv:1404.4091 [hep-th]; T. He, P. Mitra, A. P. Porfyriadis and A. Strominger, New Symmetries of Massless QED, JHEP 1410, 112 (2014) [arXiv:1407.3789 [hep-th]]; A. Strominger and A. Zhiboedov, Gravitational Mem- ory, BMS Supertranslations and Soft Theorems, arXiv:1411.5745 [hep-th]. D. Kapec, V. Lysov and A. Strominger, Asymptotic Symmetries of Massless QED in Even Di- mensions, arXiv:1412.2763 [hep-th]. D. Kapec, V. Lysov, S. Pasterski and A. Strorefsdelminger, Higher-Dimensional Supertranslations and WeinbergÕs Soft Graviton Theorem, arXiv:1502.07644 [gr-qc];  Kapec, Daniel, Raclariu, Ana-Maria and Strominger, Andrew,
  "Area, Entanglement Entropy and Supertranslations at Null Infinity,"
  arXiv:1603.07706
  Kapec, Daniel, Lysov, Vyacheslav, Pasterski, Sabrina and Strominger, Andrew,
  "Higher-Dimensional Supertranslations and Weinberg's Soft Graviton Theorem,"
  arXiv:1502.07644
  Strominger, Andrew and Zhiboedov, Alexander,
  "Gravitational Memory, BMS Supertranslations and Soft Theorems,"
  JHEP01(2016)086, arXiv:1411.5745
  Cachazo, Freddy and Strominger, Andrew,
  "Evidence for a New Soft Graviton Theorem,"
  arXiv:1404.4091
  He, Temple, Lysov, Vyacheslav, Mitra, Prahar and Strominger, Andrew,
  "BMS supertranslations and Weinberg's soft graviton theorem,"
  JHEP05(2015)151, arXiv:1401.7026
  Strominger, Andrew,
  "On BMS Invariance of Gravitational Scattering,"
  JHEP07(2014)152, arXiv:1312.2229
  S.~Weinberg, Phys.\ Rev.\ {\bf 140}, B516, 1965.
\bibitem{ckn}  A.~G.~Cohen, D.~B.~Kaplan and A.~E.~Nelson,
  ``Effective field theory, black holes, and the cosmological constant,''
  Phys.\ Rev.\ Lett.\  {\bf 82}, 4971 (1999)
  doi:10.1103/PhysRevLett.82.4971
  [hep-th/9803132].
  
  \bibitem{gidd} M.~Gary and S.~B.~Giddings,
  ``Constraints on a fine-grained AdS/CFT correspondence,''
  Phys.\ Rev.\ D {\bf 94}, no. 6, 065017 (2016)
  doi:10.1103/PhysRevD.94.065017
  [arXiv:1106.3553 [hep-th]];
  M.~Gary and S.~B.~Giddings,
  ``The Flat space S-matrix from the AdS/CFT correspondence?,''
  Phys.\ Rev.\ D {\bf 80}, 046008 (2009)
  doi:10.1103/PhysRevD.80.046008
  [arXiv:0904.3544 [hep-th]];
   M.~Gary, S.~B.~Giddings and J.~Penedones,
  ``Local bulk S-matrix elements and CFT singularities,''
  Phys.\ Rev.\ D {\bf 80}, 085005 (2009)
  doi:10.1103/PhysRevD.80.085005
  [arXiv:0903.4437 [hep-th]].
  
 
  
\end{thebibliography}
\end{document}